\documentclass[%
preprint,
amsmath,amssymb,
aps,
]{revtex4-1}

\usepackage{graphicx}
\usepackage{dcolumn}
\usepackage{bm}
\usepackage[utf8]{inputenc}
\usepackage{hyperref}
\usepackage{color}
\usepackage{epstopdf}
\usepackage{setspace}
\usepackage{geometry}
\begin{document}
	\setlength{\baselineskip}{18pt}
	
	\title {Tunable all-optical logic gates based on nonreciprocal topologically protected edge modes}
	
	\author{Jie Xu$^{1,2,7}$, Panpan He$^{3,4}$, Delong Feng$^{1,2}$, Yamei Luo$^{1,2}$, Siqiang Fan$^5$, Kangle Yong$^{1,2,8}$, Kosmas L. Tsakmakidis$^{6,9}$}
	
	\affiliation{$^1$School of Medical Information and Engineering, Southwest Medical University, Luzhou 646000, China}
	\affiliation{$^2$Medical Engineering \& Medical Informatics Integration and Transformational Medicine of Luzhou Key Laboratory, Luzhou 646000, China}
	\affiliation{$^3$Luzhou Key Laboratory of Intelligent Control and Application of Electronic Devices, Luzhou Vocational \& Technical College, Luzhou 646000, China}
	\affiliation{$^4$School of Electrical and Electronic Engineering, Luzhou Vocational \& Technical College, Luzhou 646000, China}
	\affiliation{$^5$Chongqing Key Laboratory of Photo-Electric Functional Materials, Chongqing 401331, China}
	\affiliation{$^6$ Section of Condensed Matter Physics Department of Physics National and Kapodistrian University of Athens Panepistimioupolis, Athens GR-157 84, Greece}
	
	\affiliation{$^7$ xujie011451@163.com}
	\affiliation{$^8$ Kangle@swmu.edu.cn}
	\affiliation{$^{9}$ ktsakmakidis@phys.uoa.gr}


	

	
	\begin{abstract}	
		All-optical logic gates have been studied intensively for their potential to enable broadband, low-loss, and high-speed communication. However, poor tunability has remained a key challenge in this field. In this paper, we propose a Y-shaped structure composed of Yttrium Iron Garnet (YIG) layers that can serve as tunable all-optical logic gates, including, but not limited to, OR, AND, and NOT gates, by applying external magnetic fields to magnetize the YIG layers. Our findings demonstrate that these logic gates are based on topologically protected one-way edge modes, ensuring exceptional robustness against imperfections and nonlocal effects while maintaining extremely high precision. Furthermore, the operating band of the logic gates is shown to be tunable. In addition, we introduce a straightforward and practical method for controlling and switching the logic gates between "work", "skip", and "stop" modes. These findings have important implications for the design of high-performance and precise all-optical integrated circuits.
	\end{abstract}
	
	\maketitle
	
	\section{Introduction}	
	Since the invention of the transistor in 1947, human society has experienced an unprecedented boom in electronic communications based on electrical signals to meet the needs of everyday life and scientific research \cite{riordan:Invention, Das:2021}. However, with the development of integrated circuits, transistors are becoming increasingly miniaturized, resulting in increased energy waste. Additionally, electronic communication still suffers from defects such as high error rates and cross-talk \cite{markov:Limits}. On the other hand, optical communication has advantages such as high-speed signal processing, error-free transmission \cite{Mueller:2010}, parallel computation \cite{Pierangeli:Large}, and low loss \cite{Tong:2003}, making it a potential candidate for the next-generation communication technology. In recent decades, the concept of integrated optical circuits has been introduced, greatly developed, and studied.
	
	All-optical logic gates (LGs) are an important component of integrated optical circuits and have received considerable attention in recent years, with interesting results in this field. Researchers have constructed various types of all-optical LGs, such as photonic crystal and Mach-Zehnder interferometer structures, using nonlinear processes \cite{almeida:all,wang:thermal,jandieri:realization} and/or interferometry \cite{he:topological,fu:all,gao:logic,wang:coherent}, and have implemented all basic logic operations. However, most LGs suffer from low contrast ratios (CRs), typically less than 30 dB. This is understandable because reflections are unavoidable in conventional optical LGs, and imperfections in their manufacturing affect the accuracy of the gates to some extent, particularly in nonlinearity-based LGs \cite{fushimi:all,chao:novel,Chai:Ultrafast}. In many studies on sub-wavelength all-optical LGs, researchers often neglect the impact of nonlocal effects on logical operations. While this is generally true in near-wavelength cases, non-local effects should be considered when the device's scale is subwavelength or even deep-subwavelength. In fact, the impact of nonlocal effects on nonreciprocal/one-way surface magnetoplasmons (SMPs) has been widely discussed in the past several years \cite{Hassani:Do,buddhiraju:Absence,Tsakmakidis:Topological}. SMPs are edge modes sustained in magneto-optical (MO) heterostructures, and many interesting and meaningful results, such as slow light \cite{Xu:Br,hu:slowing,hu:surface}, overcoming the time-bandwidth limit \cite{Tsakmakidis:Br}, and rainbow trapping \cite{Xu:Sl,Shen:Tr}, have been discovered. Recently, we proposed a method to implement (sub-wavelength) all-optical logic operations using one-way SMP modes \cite{Xu:Alloptical}. This type of one-way electromagnetic (EM) mode has been proven to be topologically protected \cite{Wang:Ob,Wang:To} in the microwave regime by several research groups, and no significant impact of non-local effects has been observed. Therefore, in this paper, we focus on such nonlocality-immune SMPs to study tunable LGs. Additionally, since guided wave modes have only one transmission direction, the problem of preparation process defects is well overcome, and unidirectional modes are immune to backscattering. More importantly, all-optical LGs based on unidirectional EM modes theoretically have an infinite contrast ratio, which means unparalleled accuracy.

	Note that in Ref.\cite{Xu:Alloptical}, the designed all-optical LGs relied on Yttrium Iron Garnet (YIG) with remanence. Consequently, although unidirectional SMPs-based all-optical LGs were implemented using MO heterostructures, their lack of tunability hindered their application in future integrated optical circuits. In this paper, we propose a Y-shaped structure composed of three YIG layers under different bias magnetic fields and theoretically analyze the dispersion relation in the three arms, which are all YIG-YIG heterostructures. We observe interesting phenomena, such as reverse propagation direction, and close and/or reopen one-way regions. More importantly, we discover highly tunable characteristics of the Y-shaped structure and the LGs, which are confirmed by full-wave simulation. Our proposed (subwavelength) tunable LGs have the potential to be applied in the design of high-performance and programmable integrated optical circuits.

	\section{Physical model and topologically protected SMPs}
	The Y-shaped configuration is a commonly used physical model in the field of all-optical LGs, which has been extensively studied in recent decades \cite{song:electrochemical,gao:lithographically,pan:optical,assunccao:phase,fu:all}. In Fig. 1(a), we propose a Y-shaped YIG-based model that enables tunable all-optical logic operations. The model comprises three straight arms, each containing two layers of YIG. Unlike our previous work \cite{Xu:Alloptical}, where YIG with remanence was used, all the YIG layers in this study are subjected to an external magnetic field ($\mathrm{H}_\mathrm{0}$) to further enhance the tunability of the LGs. It should be noted that metals can always be considered as perfect electric conductor (PEC) walls in the microwave regime \cite{Pozar:Mi}. For simplicity, as shown in Fig. 1(b), the arm with YIG layers having the same magnetization is referred to as 'EYYE-s', where "E" represents the PEC boundary, "Y" represents YIG, and "s" symbolizes the same magnetization direction. Similarly, the structure with YIG layers having opposite magnetization directions is labeled 'EYYE-r'. To achieve basic logic operations based on one-way modes, the key is to establish two separate one-way channels that allow efficient transfer of the EM wave/signal. The question then arises as to how to design suitable arms and how to efficiently tune the structure according to our needs.
	
	\begin{figure}[ht]
		\centering\includegraphics[width=4 in]{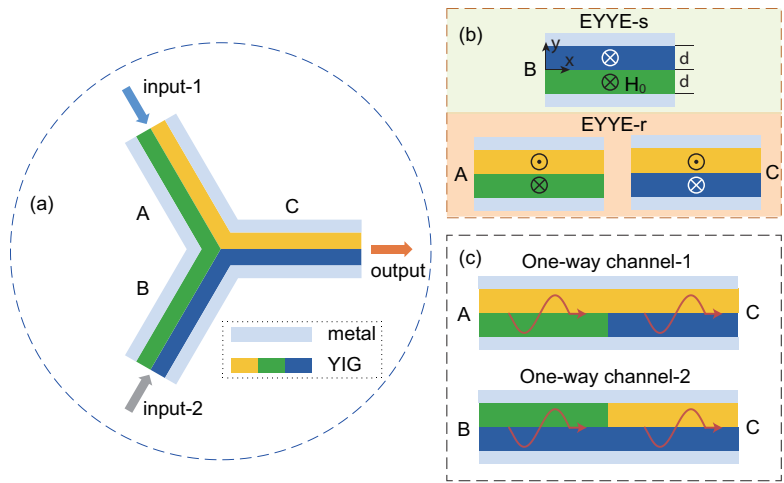}
		\caption{ (a) The schematic of the Y-shaped structure of all-optical logic operations. (b) Two types of arms are shown, i.e. the 'EYYE-s' and the 'EYYE-r'. (c) Pre-designed two one-way channels in our proposed structure. Note that, in this paper, we use $\omega_\mathrm{0}^\mathrm{a}$, $\omega_\mathrm{0}^\mathrm{b}$, and $\omega_\mathrm{0}^\mathrm{c}$ to clarify the procession angular frequencies ($\omega_\mathrm{0}$) for green-colored YIG, yellow-colored YIG and  blue-colored YIG layers, respectively.}\label{Fig1}
	\end{figure}
	
	To achieve this, one must first study the dispersion relation of the SMPs in those arms. The 'EYYE-r' contains two layers of YIG with two different relative permeability and for the lower ($\bar{\mu}_{\mathrm{a}}$) and upper ($\bar{\mu}_{\mathrm{b}}$) YIG, we have
	\begin{equation}
		\bar{\mu}_{\mathrm{a}}=\left[\begin{array}{ccc}
			\mu_1^{\mathrm{a}} & -i \mu_2^{\mathrm{a}} & 0 \\
			i \mu_2^{\mathrm{a}} & \mu_1^{\mathrm{a}} & 0 \\
			0 & 0 & 1
		\end{array}\right], \quad \bar{\mu}_{\mathrm{b}}=\left[\begin{array}{ccc}
			\mu_1^{\mathrm{b}} & i \mu_2^{\mathrm{b}} & 0 \\
			-i \mu_2^{\mathrm{b}} & \mu_1^{\mathrm{b}} & 0 \\
			0 & 0 & 1
		\end{array}\right]
	\end{equation}
	where $\mu_1 =1+\frac{\omega_\mathrm{m}\left(\omega_\mathrm{0}-i v \omega\right)}{\left(\omega_\mathrm{0}-i v \omega\right)^2-\omega^2}$ and $\mu_2 =\frac{\omega_\mathrm{m} \omega}{\left(\omega_\mathrm{0}-i v \omega\right)^2-\omega^2}$. $\omega$, $\omega_\mathrm{m}$, $\nu$ and $\omega_\mathrm{0}=\mu_\mathrm{0} \gamma \mathrm{H}_\mathrm{0}$ refer respectively to the angular frequency, the characteristic circular frequency, the damping factor, and the procession angular frequency\cite{Pozar:Mi}. Please note that the superscripts 'a' and 'b' represent the lower and upper layers, respectively. In this paper, we assume that the magnetic-field direction in the lower layer is permanently oriented in the -z direction. By applying Maxwell's equations and three boundary conditions in the 'EYYE-r' arm, one can easily calculate the dispersion relation of the SMPs sustained on the YIG-YIG interface. The dispersion relation takes the following form
	\begin{equation}
		\mu_v^\mathrm{a}\left[\frac{\mu_2^\mathrm{b}}{\mu_1^\mathrm{b}} k+\frac{\alpha_\mathrm{b}}{\tanh \left(\alpha_\mathrm{b} d\right)}\right]+\mu_v^\mathrm{b}\left[\frac{\mu_2^\mathrm{a}}{\mu_1^\mathrm{a}} k+\frac{\alpha_\mathrm{a}}{\tanh \left(\alpha_\mathrm{a} d\right)}\right]=0	
	\end{equation}
	where $\alpha_\mathrm{a}=\sqrt{k^2-\varepsilon_\mathrm{m} \mu_v^\mathrm{a} k_\mathrm{0}^2}$, $\alpha_\mathrm{b}=\sqrt{k^2-\varepsilon_\mathrm{m} \mu_v^\mathrm{b} k_\mathrm{0}^2}$, and $\mu_v=\mu_1-\mu_1^2/\mu_2$. 	
	Equation (2) reveals that the SMPs in the 'EYYE-r' arm exhibit different propagation properties for opposite wavenumbers, i.e., $k_1 = -k_2$, which is a well-known nonreciprocity effect. More importantly, adjusting the external magnetic field can create a special one-way region where the waves propagate in only one specific direction. The asymptotic frequencies (AFs) of the SMPs in the 'EYYE-r' arm can be derived and calculated from Eq. (2). We found four AFs, which can be described by the following equations:
	\begin{equation}
		\omega_{\mathrm{sp}}^{(+)}=
		\begin{aligned}
			& \begin{cases}
				&\omega_{\mathrm{sp}}^{(+1)}=\omega_\mathrm{0}^\mathrm{a}+\omega_\mathrm{m} \\
				&\omega_{\mathrm{sp}}^{(+2)}=\omega_\mathrm{0}^\mathrm{b}+\omega_\mathrm{m}
			\end{cases} \\
		\end{aligned}
	\end{equation}
	\begin{equation}
		\omega_{\mathrm{sp}}^{(-)}=
		\begin{aligned}
			& \begin{cases}
				&\omega_{\mathrm{sp}}^{(-1)}=\frac{(\omega_\mathrm{0}^\mathrm{a}+\omega_\mathrm{0}^\mathrm{b}+\omega_\mathrm{m})+\sqrt{(\omega_\mathrm{0}^\mathrm{a}+\omega_\mathrm{0}^\mathrm{b}+\omega_\mathrm{m})^2-2 (2 \omega_\mathrm{0}^\mathrm{a} \omega_\mathrm{0}^\mathrm{b}+\omega_\mathrm{0}^\mathrm{a} \omega_\mathrm{m}+\omega_\mathrm{0}^\mathrm{b} \omega_\mathrm{m})}}{2} \\
				&\omega_{\mathrm{sp}}^{(-2)}=\frac{(\omega_\mathrm{0}^\mathrm{a}+\omega_\mathrm{0}^\mathrm{b}+\omega_\mathrm{m})-\sqrt{(\omega_\mathrm{0}^\mathrm{a}+\omega_\mathrm{0}^\mathrm{b}+\omega_\mathrm{m})^2-2 (2 \omega_\mathrm{0}^\mathrm{a} \omega_\mathrm{0}^\mathrm{b}+\omega_\mathrm{0}^\mathrm{a} \omega_\mathrm{m}+\omega_\mathrm{0}^\mathrm{b} \omega_\mathrm{m})}}{2}
			\end{cases} \\
		\end{aligned}
	\end{equation}
	$\omega_\mathrm{sp}^{+}$ and $\omega_\mathrm{sp}^{-}$ indicate the AF as $k \to +\infty$ and $k \to -\infty$, respectively. In fact, the value of $\omega_\mathrm{sp}^{+}$ corresponds to the zero point of $\mu_v^\mathrm{a}$ or $\mu_v^\mathrm{b}$. Similarly, the dispersion relation of the SMPs in the 'EYYE-s' arm can be directly obtained from Eq. (2) by replacing $\mu_2^\mathrm{b}$, $\mu_1^\mathrm{b}$, $\mu_v^\mathrm{b}$, and $\alpha_\mathrm{b}$ with $-\mu_2^\mathrm{c}$, $\mu_1^\mathrm{c}$, $\mu_v^\mathrm{c}$, and $\alpha_\mathrm{c}$, respectively. In this case, the permeability ($\bar{\mu_\mathrm{c}}$) of the upper YIG has the same form as $\bar{\mu_\mathrm{a}}$, and the corresponding dispersion equation can be written as follows:
	\begin{equation}
		\mu_v^\mathrm{a}\left[\frac{-\mu_2^\mathrm{c}}{\mu_1^\mathrm{c}} k+\frac{\alpha_\mathrm{c}}{\tanh \left(\alpha_\mathrm{c} d\right)}\right]+\mu_v^\mathrm{c}\left[\frac{\mu_2^\mathrm{a}}{\mu_1^\mathrm{a}} k+\frac{\alpha_\mathrm{a}}{\tanh \left(\alpha_\mathrm{a} d\right)}\right]=0	
	\end{equation}
	There are also four potential AFs in the 'EYYE-s' arm, which have the following form:
	\begin{equation}
		\omega_{\mathrm{sp}}^{(+)}=
		\begin{aligned}
			& \begin{cases}
				&\omega_{\mathrm{sp}}^{(+1)}=\omega_\mathrm{0}^\mathrm{a}+\omega_\mathrm{m} \\
				&\omega_{\mathrm{sp}}^{(+2)}=\frac{(\omega_\mathrm{0}^\mathrm{c}-\omega_\mathrm{0}^\mathrm{a})+\sqrt{(\omega_\mathrm{0}^\mathrm{c}-\omega_\mathrm{0}^\mathrm{a})^2+2 (2 \omega_\mathrm{0}^\mathrm{a} \omega_\mathrm{0}^\mathrm{c}+\omega_\mathrm{0}^\mathrm{a} \omega_\mathrm{m}+\omega_\mathrm{0}^\mathrm{c} \omega_\mathrm{m})}}{2}
			\end{cases} \\
		\end{aligned}
	\end{equation}
	\begin{equation}
		\omega_{\mathrm{sp}}^{(-)}=
		\begin{aligned}
			& \begin{cases}
				&\omega_{\mathrm{sp}}^{(-1)}=\omega_\mathrm{0}^\mathrm{c}+\omega_\mathrm{m}\\
				&\omega_{\mathrm{sp}}^{(-2)}=\frac{(\omega_\mathrm{0}^\mathrm{a}-\omega_\mathrm{0}^\mathrm{c})+\sqrt{(\omega_\mathrm{0}^\mathrm{c}-\omega_\mathrm{0}^\mathrm{a})^2+2 (2 \omega_\mathrm{0}^\mathrm{a} \omega_\mathrm{0}^\mathrm{c}+\omega_\mathrm{0}^\mathrm{a} \omega_\mathrm{m}+\omega_\mathrm{0}^\mathrm{c} \omega_\mathrm{m})}}{2}
			\end{cases} \\
		\end{aligned}
	\end{equation}
	
	\begin{figure}[pt]
		\centering\includegraphics[width=5 in]{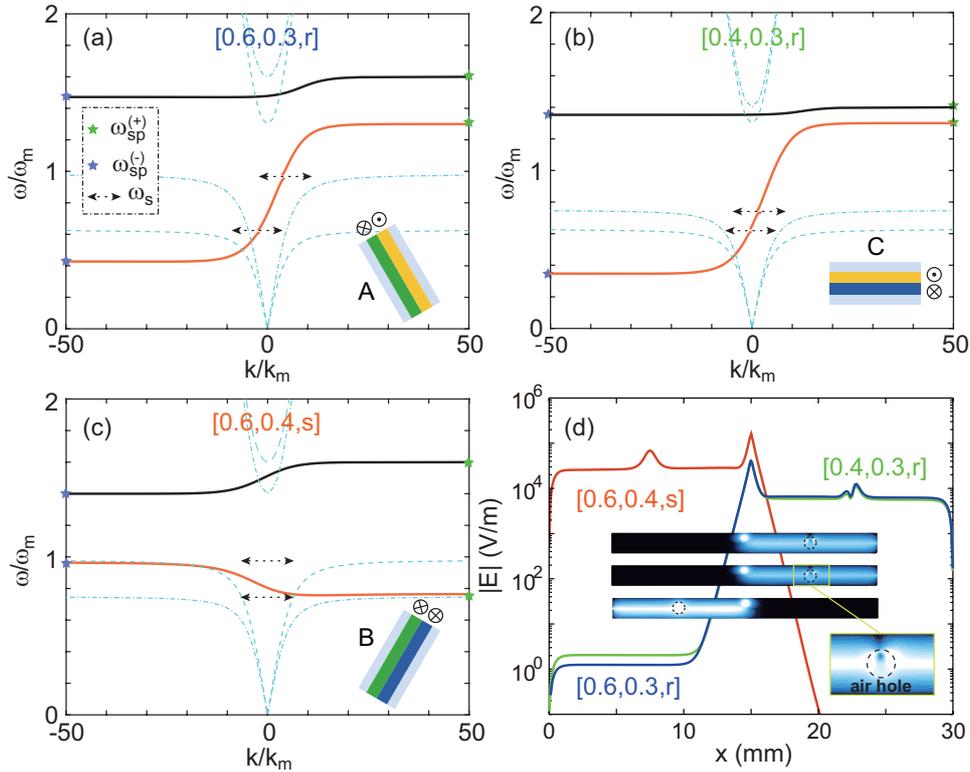}
		\caption{ (a-c) The dispersion diagrams of three arms are shown, in which the lower YIG has $\omega_\mathrm{0}$ values of $0.6\omega_\mathrm{m}$, $0.4\omega_\mathrm{m}$, and $0.6\omega_\mathrm{m}$, respectively, while the upper YIG has $\omega_\mathrm{0}$ values of $0.3\omega_\mathrm{m}$, $0.3\omega_\mathrm{m}$, and $0.4\omega_\mathrm{m}$", respectively. Note that 'r' and 's' indicate the 'EYYE-r' arm and 'EYYE-s' arm, respectively, and the magnetization orientation of the lower YIG is permanently -z. The cyan lines represent the edge of the bulk zones, while the black arrows indicate the location of $\omega = \omega_\mathrm{s}$. Stars show the corresponding asymptotic frequencies in each case. (d) The simulated electric field distributions of the three cases are shown for $f=0.8f_\mathrm{m}$. The other parameters are (a-c) $d=0.02 \lambda_\mathrm{m}$, $\nu = 0$ and (d) $\nu = 0.001 \omega_\mathrm{m}$.}\label{Fig2}
	\end{figure}
	Based on Eqs. (4) and (5), we plot the dispersion curves for the SMPs in both the 'EYYE-r' and 'EYYE-s' arms as $d=0.02 \lambda_\mathrm{m}$ ($\lambda_\mathrm{m}=2\pi c/\omega_\mathrm{m}=c/f_\mathrm{m}$) and $\nu = 0$ (lossless condition). Three different values of $\omega_\mathrm{0}$ ($\mathrm{H}_\mathrm{0}$) are applied in the three arms, and for convenience, we introduce a simple notation - '$\left[\alpha, \beta, \theta \right]$' - in which $\alpha$ and $\beta$ represent the absolute values of the normalized $\omega_\mathrm{0}$ ($\bar{\omega_\mathrm{0}}=\omega_\mathrm{0}/\omega_\mathrm{m}$) for the lower and upper YIG, while $\theta$ could be either 'r' referring to the 'EYYE-r' arm or 's' referring to the 'EYYE-s' arm. For example, [0.6, 0.3, r] in Fig. 2(a) implies that the dispersion curve is calculated in the 'EYYE-r' arm, where $\omega_\mathrm{0}^\mathrm{a}=0.6\omega_\mathrm{m}$ and $\omega_\mathrm{0}^\mathrm{b}=0.3\omega_\mathrm{m}$. In Fig. 2(a), the green and blue stars represent $\omega_{\mathrm{sp}}^{(+)}$ and $\omega_{\mathrm{sp}}^{(-)}$, respectively. The red and black lines indicate the dispersion curves of SMPs on the YIG-YIG interface, and due to the deep-subwavelength thickness of the YIG layers, the bulk zones are significantly compressed \cite{Xu:Sl, Zhou:realization}. Therefore, it is believed that almost all the SMPs on the red and black lines are one-way EM modes except for the SMPs located near the resonant frequencies of YIG ($\omega_\mathrm{s}=\sqrt{\omega_\mathrm{0}^2+\omega_\mathrm{0}\omega_\mathrm{m}}$), which are marked by black arrows. As depicted in the inset of Fig. 2(a), the case of [0.6, 0.3, r] can be treated as one of the input arms (arm 'A') of the Y-shaped heterostructure. We also calculate the dispersion curves for the other arms in Figs. 2(b) and 2(c). As a result, similar to the first case, there are two one-way regions in both cases. However, in the case of [0.6, 0.4, s], the EM waves within the lower one-way region have negative group velocities ($v_\mathrm{g}<0$). 
	
	Based on Eqs. (3), (4), (6), and (7) as mentioned earlier, the one-way regions are defined by the AFs (green and blue stars in Fig. 2). For the three cases discussed above, the regions are: (a) [0.428$f_\mathrm{m}$, 1.3$f_\mathrm{m}$] and [1.472$f_\mathrm{m}$, 1.6$f_\mathrm{m}$], (b) [0.3475$f_\mathrm{m}$, 1.3$f_\mathrm{m}$] and [1.3525$f_\mathrm{m}$, 1.4$f_\mathrm{m}$], and (c) [0.766$f_\mathrm{m}$, 0.966$f_\mathrm{m}$] and [1.4$f_\mathrm{m}$, 1.6$f_\mathrm{m}$]. Therefore, to design two one-way channels, the frequencies used must be located within the [0.766$f_\mathrm{m}$, 0.966$f_\mathrm{m}$] region (the red line region in Fig. 2(c)).
	In addition, the loss effect and the robustness of the one-way propagation of SMPs are examined using full-wave simulations, as illustrated in Fig. 2(d). In this case, we consider $\nu=0.001\omega_\mathrm{m}$ and $f=0.8f_\mathrm{m}$, and air holes with a radius of $r=0.5$ mm ($\sim 0.008\lambda_\mathrm{m}$) are placed on the YIG-YIG interface. 
	The simulation results show good agreement with the theoretical analysis, and the imperfections have a negligible impact on the one-way SMPs. 
	It is also worth noting that recent studies have questioned the robustness of one-way SMPs, with the nonlocal effect being a major focus of these works\cite{Hassani:Do,buddhiraju:Absence}. Here, we emphasize that the one-way SMPs studied in this paper are theoretically topologically protected, which has been theoretically demonstrated\cite{silveirinha:chern,Wang:Re} and experimentally proved\cite{holmes:Experimental, Wang:Ob} by many groups. This nonlocality-immune property is particularly evident in cases where the waveguide is relatively thick or the wavenumber (k) is relatively small\cite{Tsakmakidis:Topological}. Our proposed logic gates in this paper are believed to be largely unaffected by nonlocal effects, given the tunability of the SMPs, which will be discussed in the next subsection.

	\section{Tunable all-optical logic gates}
	In our theoretical analysis, we have shown that a Y-shaped structure consisting of magnetized YIG layers can support two independent one-way channels, making it suitable as a logical gate\cite{Xu:Alloptical}. More importantly, benefiting from the tunability of the topologically protected one-way SMPs, the proposed LGs should be easily tunable by changing the bias magnetic fields. In the following sections, we demonstrate the tunability of our proposed logical gates in detail. 
	Firstly, we study the impact of $\mathrm{H}_\mathrm{0}$ on AFs, which always define the one-way regions. As displayed in Fig. 3(a), four AFs in the 'EYYE-r' arm 'A' ('C'), are plotted as a function of $\omega_\mathrm{0}^\mathrm{a}$ ($\omega_\mathrm{0}^\mathrm{c}$) and $\omega_\mathrm{0}^\mathrm{b}$. Figure 3(b) depicts the similar relationship between AFs and $\omega_\mathrm{a}$ and $\omega_\mathrm{c}$. 
	To differentiate between the four distinct AFs in Eqs. (3, 4) ('EYYE-r') and (6, 7) ('EYYE-s'), we use the names $\omega_\mathrm{sp}^{(+1)}$, $\omega_\mathrm{sp}^{(+2)}$, $\omega_\mathrm{sp}^{(-1)}$, and $\omega_\mathrm{sp}^{(-2)}$. 
	Notably, as $\omega_\mathrm{0}$ ($\mathrm{H}_\mathrm{0}$) changes, the values of AFs and their numerical relationships may also change. This can lead to a reversal of the group velocity and the transmission direction of EM signals in LGs. Therefore, any changes in the AFs can affect the functionality of the LGs.
	\begin{figure}[pt]
		\centering\includegraphics[width=6 in]{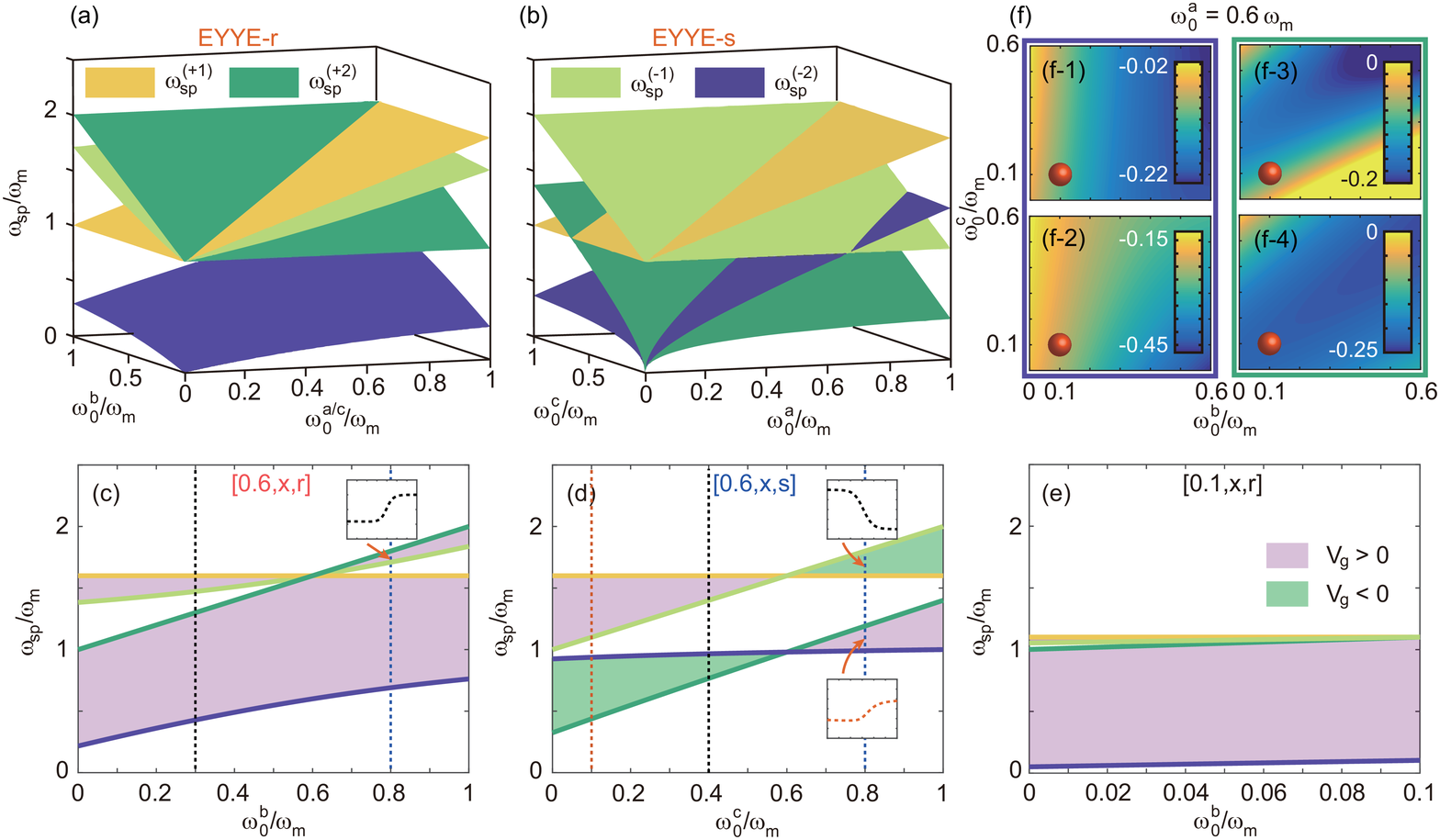}
		\caption{(a,b) The asymptotic frequencies (AFs) are plotted as a function of $\omega_\mathrm{0}^\mathrm{a}$ (or $\omega_\mathrm{0}^\mathrm{c}$) and $\omega_\mathrm{0}^\mathrm{b}$ for (a) the 'EYYE-r' arm and (b) the 'EYYE-s' arm. (c-e) AFs are plotted as a function of $\omega_\mathrm{0}^\mathrm{b}$ when (c,d) $\omega_\mathrm{0}^\mathrm{a}=0.6 \omega_\mathrm{m}$ and (e) $\omega_\mathrm{0}^\mathrm{c}=0.1 \omega_\mathrm{m}$. (f) Four constructed equations ($y_1$, $y_2$, $y_3$, and $y_4$) as shown in Eq. (8) are plotted as functions of $\omega_\mathrm{0}^\mathrm{b}$ and $\omega_\mathrm{0}^\mathrm{c}$ when $\omega_\mathrm{0}^\mathrm{a}=0.6\omega_\mathrm{m}$.}\label{Fig3}
	\end{figure}
	
	To illustrate the changes in AFs and one-way regions, we set the lower YIG $\omega_\mathrm{0}$ to $0.6\omega_\mathrm{m}$ and assume $0<\omega_\mathrm{0}<\omega_\mathrm{m}$ for the upper YIG in both 'EYYE-r' (Fig. 3(c)) and 'EYYE-s' (Fig. 3(d)) arms. As $\omega_\mathrm{0}$ ($\omega_\mathrm{0}^\mathrm{b}$) of the upper YIG varies from 0 to $0.6\omega_\mathrm{m}$, the lower one-way region gradually widens, while the upper one-way region becomes smaller and eventually closes at $\omega_\mathrm{0}^\mathrm{b}=0.6\omega_\mathrm{m}$. The black dashed line represents the [0.6, 0.3, r] case discussed earlier in Fig. 2(a), in which two clear one-way regions are present (excluding the local area near $\omega=\omega_\mathrm{s}$). As $\omega_\mathrm{0}^\mathrm{b}$ increases further, for the 'EYYE-r' arm, the first one-way region is compressed slightly, while a new one-way region bounded by $\omega_{\mathrm{sp}}^{(-1)}$ and $\omega_{\mathrm{sp}}^{(+2)}$ emerges with a forward propagation direction ($v_\mathrm{g}>0$).The inset of Fig. 3(c) displays a zoomed-in dispersion curve for the case of $\omega_\mathrm{0}^\mathrm{b}=0.8\omega_\mathrm{m}$ (blue line). In contrast, the 'EYYE-s' arm behaves differently. As shown in Fig. 3(d), when $\omega_\mathrm{0}^\mathrm{c}$ (in the upper YIG) is increased, the propagation direction of SMPs in the lower one-way region changes from backward ($v_\mathrm{g}<0$) to forward ($v_\mathrm{g}>0$), and the one-way region closes and reopens. Similar phenomena of reversing propagation direction and close-reopen one-way regions are observed in the higher regime as well. The black and blue dashed lines in Fig. 3(d) indicate cases where $\omega_\mathrm{0}^\mathrm{c}=0.4\omega_\mathrm{m}$ and $\omega_\mathrm{0}^\mathrm{c}=0.8\omega_\mathrm{m}$, respectively, with $\omega_\mathrm{0}^\mathrm{a}=0.6\omega_\mathrm{m}$ in both cases. The insets in Fig. 3(d) show the reversed one-way regions and the dispersion curves of SMPs.
	
	The question of whether the group velocity will reverse in one-way systems can be answered by determining if the system's symmetry or chirality is broken. As demonstrated in Figure 3(c), within the lower one-way region of the 'EYYE-r' arm, the SMPs can propagate only in the forward direction, regardless of which layer has a higher $\mathrm{H}_\mathrm{0}$ ($\omega_\mathrm{0}$). We consider two conditions, [0.6, 0.4, r] and [0.4, 0.6, r], where the propagation directions are the same. This is because the second case can be treated as the entire system of the first case revolving 180 degrees around the propagation direction, and thus the system's symmetry/chirality is conserved. In contrast, [0.6, 0.4, s] and [0.4, 0.6, s] have opposite propagation directions because they cannot be obtained by simply rotating each other, and thus the system's symmetry/chirality is broken when changing $\omega_\mathrm{0}$ accordingly.
	
	To achieve a relatively broad one-way band, it is necessary that $\omega_\mathrm{0}^{c}$ in arm 'B' is small enough, as shown in Figure 3(d). Thus, we select $\omega_\mathrm{0}^\mathrm{c} = 0.1\omega_\mathrm{m}$ (marked by the red dashed line in Figure 3(d)), and Figure 3(e) depicts the corresponding AFs and one-way regions as functions of $\omega_\mathrm{0}^\mathrm{b}$. For this case, $\omega_{\mathrm{sp}}^{(-2)} \simeq 0.048\omega_\mathrm{m}$ and $\omega_{\mathrm{sp}}^{(+2)} =\omega_\mathrm{m}+ \omega_\mathrm{b}$. With $\omega_\mathrm{0}^\mathrm{a}=0.6\omega_\mathrm{m}$ and $\omega_\mathrm{0}^\mathrm{c}=0.1\omega_\mathrm{m}$ fixed, the only remaining unknown parameter in the Y-shaped structure is $\omega_\mathrm{0}^\mathrm{b}$. Ideally, we aim for the whole one-way region with $v_\mathrm{g}<0$ in arm 'B' to be the working band of the LGs. Based on our calculations, we can achieve this goal if $0<\omega_\mathrm{0}^\mathrm{b}<0.31\omega_\mathrm{m}$. However, in most cases, to ensure that the entire one-way region of arm 'B' is the working band of LGs, we need to ensure that $\omega_{\mathrm{sp}}^{(-2)}$ (the blue line in Figure 3(d)) and $\omega_{\mathrm{sp}}^{(+2)}$ (the green line in Figure 3(d)) are both inside the one-way regions with $v_\mathrm{g}>0$ in arms 'A' and 'C'. To accomplish this, we construct the following equations:
	\begin{equation}
		\begin{aligned}
			& \begin{cases}
				&y_1 = (\textcolor{black}{\omega_{\mathrm{sp\_B}}^{(-2)}}-\textcolor{black}{\omega_{\mathrm{sp\_A}}^{(-2)}})(\textcolor{black}{\omega_{\mathrm{sp\_B}}^{(-2)}}-\textcolor{black}{\omega_{\mathrm{sp\_A}}^{(+2)}}) \\	
				&y_2 = (\textcolor{black}{\omega_{\mathrm{sp\_B}}^{(-2)}}-\textcolor{black}{\omega_{\mathrm{sp\_C}}^{(-2)}})(\textcolor{black}{\omega_{\mathrm{sp\_B}}^{(-2)}}-\textcolor{black}{\omega_{\mathrm{sp\_C}}^{(+2)}}) \\	
				&y_3 = (\textcolor{black}{\omega_{\mathrm{sp\_B}}^{(+2)}}-\textcolor{black}{\omega_{\mathrm{sp\_A}}^{(-2)}})(\textcolor{black}{\omega_{\mathrm{sp\_B}}^{(-2)}}-\textcolor{black}{\omega_{\mathrm{sp\_A}}^{(+2)}}) \\	
				&y_4 = (\textcolor{black}{\omega_{\mathrm{sp\_B}}^{(+2)}}-\textcolor{black}{\omega_{\mathrm{sp\_C}}^{(-2)}})(\textcolor{black}{\omega_{\mathrm{sp\_B}}^{(-2)}}-\textcolor{black}{\omega_{\mathrm{sp\_C}}^{(+2)}}) \\				
			\end{cases} \\
		\end{aligned}
	\end{equation}
	where 'A/B/C' represent arm 'A'/'B'/'C'. In this context, it is worth noting that arms 'A' and 'C' belong to the 'EYYE-r' type, while arm 'B' belongs to the 'EYYE-s' type. The AFs are represented by $\omega_{\mathrm{sp}}^{(+)}$ and $\omega_{\mathrm{sp}}^{(-)}$, which are given by Eqs. (3), (4), (6), and (7). Equation (8) determines whether $\omega_{\mathrm{sp}}^{(-2)}$ in arm 'B' lies within the one-way region of arm 'A', which occurs for $y_1<0$. If $y_1<0$, $y_2<0$, $y_3<0$, and $y_4<0$ at the same time, it means that the entire one-way region with $v_\mathrm{g}<0$ in arm 'B' lies within the one-way regions of both arms 'A' and 'C'. Figure 3(f) represents the functions of $y_1$ ((f-1)), $y_2$ ((f-2)), $y_3$ ((f-3)), and $y_4$ ((f-4)) based on $\omega_\mathrm{0}^\mathrm{b}$ and $\omega_\mathrm{0}^\mathrm{c}$ when $\omega_\mathrm{0}^\mathrm{a} = 0.6\omega_\mathrm{m}$. We observe that $y_1$, $y_2$, and $y_4$ are always negative, while $y_3$ can be positive for relatively large $\omega_\mathrm{0}^\mathrm{b}$ and small $\omega_\mathrm{0}^\mathrm{c}$. Therefore, we set $\omega_\mathrm{0}^\mathrm{a} = 0.6\omega_\mathrm{m}$ and $\omega_\mathrm{0}^\mathrm{c} = 0.1\omega_\mathrm{m}$, and keep $\omega_\mathrm{0}^\mathrm{b}$ relatively small, such as $\omega_\mathrm{0}^\mathrm{b} = 0.1\omega_\mathrm{m}$ (marked by red balls in Fig. 3(f)), to ensure that the entire one-way region in arm 'B' corresponds to the working band of LGs.
	
	Figures 4(a)-4(c) present dispersion curves for the scenario where $\omega_\mathrm{0}^\mathrm{a} = 0.6\omega_\mathrm{m}$ and $\omega_\mathrm{0}^\mathrm{b}=\omega_\mathrm{0}^\mathrm{c} = 0.1\omega_\mathrm{m}$. Similar to Fig. 2, there is a one-way region with $v_\mathrm{g}>0$ in arm 'A' (depicted as a red-line region in Fig. 4(a)) and arm 'C' (depicted as a red-line region in Fig. 4(b)). Additionally, there is a one-way region with $v_\mathrm{g}<0$ in arm 'B' (depicted as a red-line region in Fig. 4(c)). Moreover, the backward one-way region is much larger than that illustrated in Fig. 2(c). Consequently, the working band of the LGs in this situation should be significantly broader. Figs. 4(d) and 4(f) show the coupling effect between arms when $f=0.8f_\mathrm{m}$, which falls within the one-way regions of interest. Consequently, two one-way channels ('A-C' and 'B-C') are established, while the EM signal cannot propagate from arm 'A' to arm 'B'. It should be noted that the first part ([0.1, 0.6, s]) of the 'B-C' channel differs from that of the 'A-B' channel ([0.6, 0.1, s]) due to the geometrical relationship between the arms. As per symmetry, the SMPs in the [0.1, 0.6, s] and [0.6, 0.1, s] structures must have opposite propagation directions. In the simulations, the EM signal can transfer efficiently in the one-way channels, while the forward transferring signal halts at the interface of arm 'A' and arm 'B'.
	
	\begin{figure}[ht]
		\centering\includegraphics[width=5 in]{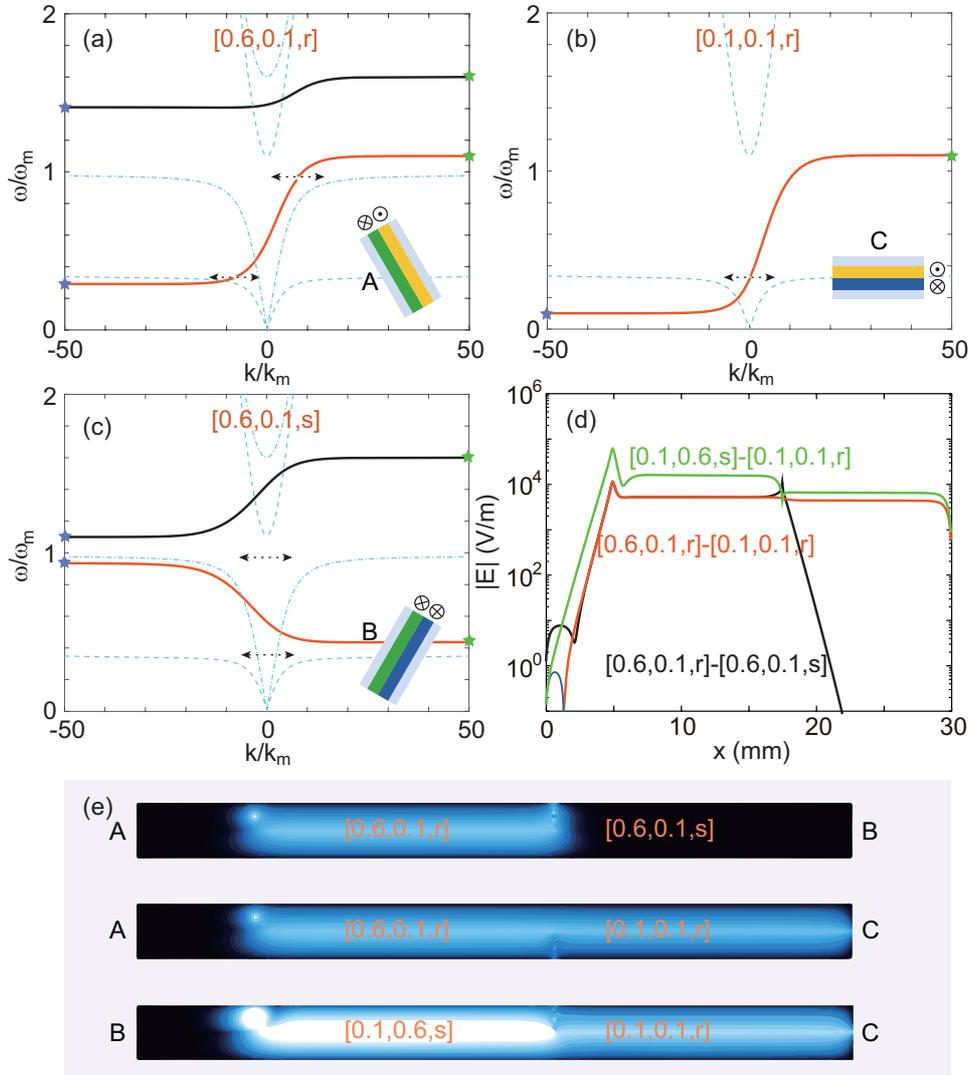}
		\caption{ (a,b) Dispersion diagrams of three arms with optimized parameters, $\omega_\mathrm{0}^\mathrm{a}=0.6\omega_\mathrm{m}$, $\omega_\mathrm{0}^\mathrm{b}=0.1\omega_\mathrm{m}$ and $\omega_\mathrm{0}^\mathrm{c}=0.1\omega_\mathrm{m}$. (d,e) The simulated electric field distribution obtained from coupling simulations containing two arms, with each arm being either the 'EYYE-r' type or 'EYYE-s' type.}\label{Fig4}
	\end{figure}

	\section{Realization of basic logic gates}
	\begin{figure}[pt]
		\centering\includegraphics[width=5 in]{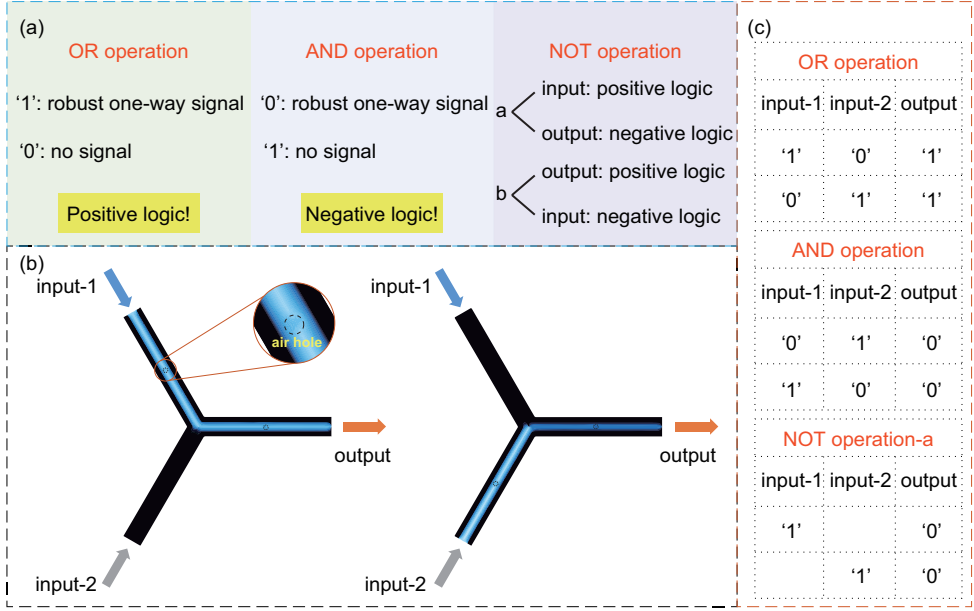}
		\caption{ (a) Theory of all-optical logic operation using the positive and/or negative logic. (b) Numerical simulations in the Y-shaped module as $f=0.8f_\mathrm{m}$, and air holes with $r=0.5$ mm were set on the YIG-YIG interfaces to verify the robustness of logic operations. (c) The truth tables of the OR, AND, and NOT operations.}\label{Fig5}
	\end{figure}
	The Y-shaped structure that is designed can function within the one-way region and perform as basic logic gates, including, but not limited to, OR, AND, and NOT gates, as presented in Fig. 5. During logical operations, arms 'A' and 'B' are considered as two input ports, with arm 'C' regarded as the output port. Primarily, the structure operates as a natural OR gate, where any input one-way EM signal can and must propagate to the output port. If we consider the presence of the EM signal as logic '1' and the absence of the EM signal as logic '0,' i.e., positive logic, then the Y-shaped structure functions as a broad OR gate that can be adjusted by external magnetic fields. Negative logic (where the presence of the EM signal is recognized as logic '0') is used for the AND gate and NOT gate. In the AND operation, any input EM signal is treated as logic '0,' resulting in the output EM signal also being logic '0.' However, the NOT operation employs negative logic in either the input or output port, with positive logic used in the remaining port. Figure 5(b) depicts simulations for the Y-shaped structure with $f=0.8f_\mathrm{m}$ when the EM signal is excited in only one of the input ports. Air holes with $r=0.5$ mm were set as imperfections on the YIG-YIG interface to show the robust function of our proposed LGs. As a result, the LGs work fine and perform extremely high CR which is larger than 200 dB (infinity in theory). Figure 5(c) shows the corresponding truth tables of the OR, AND, and NOT operations.
	
	The Y-shaped structure is designed to function within the one-way region and can serve as basic logic gates, including but not limited to OR, AND, and NOT gates, as illustrated in Fig. 5. During logical operations, arms 'A' and 'B' are the input ports, while arm 'C' is the output port. The structure operates as a natural OR gate, where any input one-way EM signal must propagate to the output port. If we assume that the presence of the EM signal is logic '1' and its absence is logic '0' (positive logic), then the Y-shaped structure functions as a versatile OR gate that can be externally adjusted by magnetic fields. However, the AND and NOT gates use negative logic, where the presence of the EM signal is recognized as logic '0.' In the AND operation, any input EM signal is treated as logic '0,' resulting in the output EM signal also being logic '0.' On the other hand, the NOT operation employs negative logic in either the input or output port, with positive logic used in the remaining port. Figure 5(b) shows simulations of the Y-shaped structure with $f = 0.8f_\mathrm{m}$ when the EM signal is excited in only one of the input ports. Imperfections on the YIG-YIG interface were introduced as air holes with $r = 0.5$ mm to demonstrate the robustness of our proposed LGs. As a result, the LGs exhibits high CR of over 200 dB (infinity in theory). Figure 5(c) presents the corresponding truth tables for the OR, AND, and NOT operations. 
	\begin{figure}[pt]
		\centering\includegraphics[width=6 in]{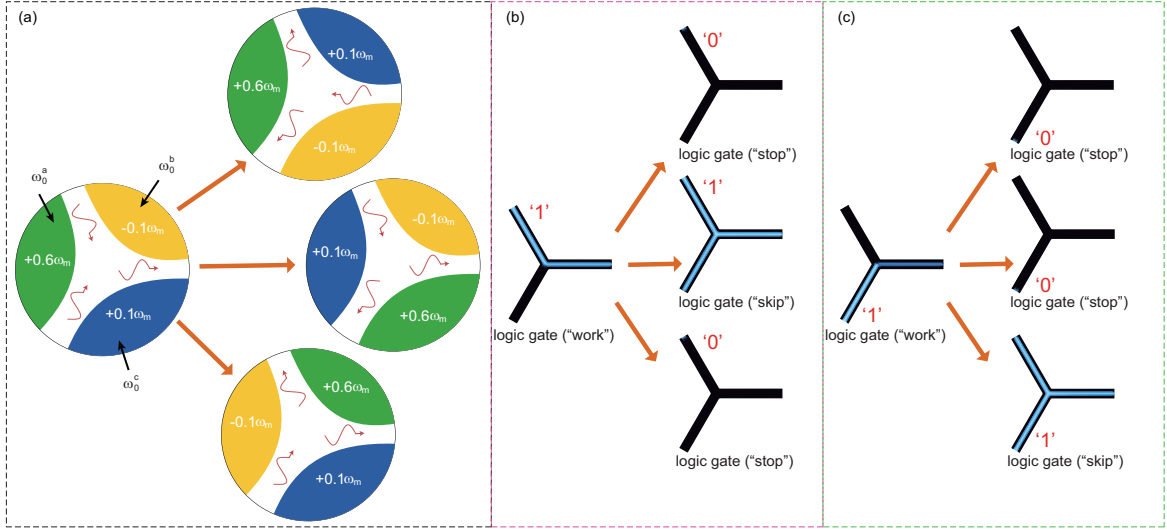}
		\caption{ (a) Switch theory for tunable LGs. The red arrow refers to the electromagnetic wave path that allows passage. (b,c) Simulation for tunable LGs by switching the external magnetic fields for input signals (b) ['1', '0'] and (c) ['0', '1']. Input '1' could be alternatively programmed to '0' ("stop" mode) or '1' ("skip" or "work" mode.)}\label{Fig6}
	\end{figure}
	
	As mentioned earlier in Fig. 3, we can control the LGs based on the Y-shaped structure with external magnetic fields ($\mathrm{H}_\mathrm{0}$). By changing the direction(s) of $\mathrm{H}_\mathrm{0}$, we can reverse the propagation direction of the one-way SMPs, as illustrated by the lower one-way regions in Fig. 3(c,d), such as [0.6, 0.4, r] transitioning to [0.6, 0.4, s]. Another notable case is that changing the direction(s) of $\mathrm{H}_\mathrm{0}$ can cause the previous one-way region to close, such as [0.6, 0.6, r] shifting towards [0.6, 0.6, s]. In this scenario, there are no one-way regions, and the entire band becomes a band gap, preventing the propagation of EM signals. Additionally, altering the value of $\mathrm{H}_\mathrm{0}$, either by increasing or decreasing it, can significantly affect the logic operations. Therefore, the operating band of our proposed LGs can be easily tunned by changing the external magnetic fields. 
	
	Besides, we suggest an innovative approach to tune LGs, as demonstrated in Fig. 6. This method achieves three modes of LGs by switching $\mathrm{H}_\mathrm{0}$ or $\omega_\mathrm{0}$, namely the "work," "stop," and "skip" modes. As initial magnetic-field parameters, we set $\omega_\mathrm{0}^\mathrm{a}=0.6\omega_\mathrm{m}$, $\omega_\mathrm{0}^\mathrm{b}=-0.1\omega_\mathrm{m}$, and $\omega_\mathrm{0}^\mathrm{c}=0.1\omega_\mathrm{m}$. It is noteworthy that the '-' sign indicates the external magnetic field is in the +z direction. Upon exchanging $\omega_\mathrm{0}^\mathrm{b}$ ($\mathrm{H}_\mathrm{0}^\mathrm{b}$) and $\omega_\mathrm{0}^\mathrm{c}$ ($\mathrm{H}_\mathrm{0}^\mathrm{c}$), it is easy to calculate that SMPs with $f=0.8f_\mathrm{m}$ have opposite propagation directions in the original arms. Similarly, exchanging $\omega_\mathrm{0}^\mathrm{c}$ ($\mathrm{H}_\mathrm{0}^\mathrm{c}$) and $\omega_\mathrm{0}^\mathrm{a}$ ($\mathrm{H}_\mathrm{0}^\mathrm{a}$) reverses the propagation direction of SMPs in arm 'B,' whereas the direction remains unchanged in other arms. Moreover, exchanging $\omega_\mathrm{0}^\mathrm{a}$ ($\mathrm{H}_\mathrm{0}^\mathrm{a}$) and $\omega_\mathrm{0}^\mathrm{b}$ ($\mathrm{H}_\mathrm{0}^\mathrm{b}$) reverses the propagation direction of SMPs in arm 'A,' whereas the direction remains unaltered in other arms. The simulation of the three modes of LG is illustrated in Figs. 6(b) and 6(c). The original mode is designated the "work" mode since it can work as LGs. Two methods can achieve the "stop" mode with EM signals being halted, and one method can accomplish the "skip" mode with EM signals skipping the present calculation. Therefore, our proposed Y-shaped LGs offers rich manipulation possibilities and is promising for programmable optical communication/devices. In contrast, traditional all-optical LGs typically operate at fixed frequencies and can be challenging to tune.

	\section{Conclusion}
	In summary, we have devised a Y-shaped structure made of YIG layers with distinct magnetizations. The arms of the structure were categorized into two types: the 'EYYE-r' type with opposing magnetization directions and the 'EYYE-s' type with identical magnetization directions. Our theoretical analysis of the 'EYYE-r' and 'EYYE-s' arms led to the construction of two one-way channels capable of supporting topologically protected one-way SMPs. Furthermore, the implementation of basic logic gates, such as OR, AND, and NOT gates, is achieved through these broadband and topological one-way SMPs, resulting in highly robust (resistant to backscattering and imperfections) and precise (theoretically infinite contrast ratio) LGs. In addition, we explored the tunability of these LGs. By adjusting external magnetic fields, the one-way region can be easily modulated, either broadened or narrowed, the propagation directions of the SMPs within the region can be completely reversed, or the region can be closed. Given the intriguing tunability of the operating band of the Y-shaped LGs. In addition, we proposed a potential application for the structure/LGs: by switching external magnetic fields, three switchable modes ("work", "skip", and "stop") can be achieved. Our proposed LGs, based on magnetized YIG, may pave the way for tunable all-optical logic operations and hold promise for high-efficiency programmable optical communication circuits.
	
	\section*{Acknowledgement}
	This work was supported by the National Natural Science Foundation of Sichuan Province (No. 2023NSFSC1309), and the open fund of Luzhou Key Laboratory of Intelligent Control and Application of Electronic Devices (No. ZK202210), Sichuan Science and Technology Program (No. 2022YFS0616), the Science and Technology Strategic Cooperation Programs of Luzhou Municipal People’s Government and Southwest Medical University (No. 2019LZXNYDJ18). J.X., K.Y., and Y.L. thanks for the support of the Innovation Laboratory of Advanced Medical Material \& Physical Diagnosis and Treatment Technology. K.L.T. was supported by the General Secretariat for Research and Technology (GSRT) and the Hellenic Foundation for Research and Innovation (HFRI) under Grant No. 4509.

	
	\bibliographystyle{unsrt}
	\bibliography{mybib}
	

\end{document}